\documentclass[prb,nobalancelastpage,twocolumn,superscriptaddress,nofootinbib]{revtex4-1}
\usepackage{color,amsthm,amsmath,amsfonts,amssymb,graphicx,bm}
\usepackage{ulem}
\usepackage{dcolumn}

\begin{document}
\title{Microscopic Theory of Polariton Lasing via Vibronically Assisted Scattering}

\author{L. Mazza}
\email{leonardo.mazza@sns.it}
\affiliation{Scuola Normale Superiore, Piazza dei Cavalieri 7, I-56126, Pisa, Italy}

\author{S. K\'ena-Cohen}
\affiliation{The Blackett Laboratory, Department of Physics, Imperial College London, London SW7 2AZ, United Kingdom}

\author{P. Michetti}
\affiliation{Institute for Theoretical Physics and Astrophysics, University of W\"urzburg, D-97074 W\"urzburg, Germany}

\author{G. C. La Rocca}
\affiliation{Scuola Normale Superiore and CNISM, Piazza dei
Cavalieri 7, I-56126, Pisa, Italy}

\begin{abstract}
Polariton lasing has recently been observed in strongly coupled
crystalline anthracene microcavities. A simple model is developed
describing the onset of the non-linear threshold based on a master
equation including the relevant relaxation processes and employing
realistic material parameters. The mechanism governing the
build-up of the polariton population - namely bosonic stimulated
scattering from the exciton reservoir via a vibronically assisted
process - is characterized and its efficiency calculated
on the basis of a microscopic theory. The role of
polariton-polariton bimolecular quenching is identified and
temperature dependent effects are discussed.
\end{abstract}

\maketitle

\section{Introduction}

In strongly coupled semiconductor microcavities~\cite{kav,deng}
the cavity mode and the excitonic resonance mix and form new
bosonic quasiparticles, the polaritons. 
Their properties differ significantly 
from those of the bare uncoupled excitations,
though they originate from them. 
The lower polariton (LP) has a peculiar dispersion law
with a deep minimum at small wavevectors, characterized by a tiny mass. 
At high densities, the build-up of a large population at the bottom of this
branch is favoured by bosonic final-state stimulation as soon as
the occupation per mode of the lower polariton states exceeds unity.
Coherent light-emission, called polariton lasing, results from this macroscopic population due to the finite lifetime of the polaritons which leak out of the cavity via their photonic component.
This is only one of the many outstanding
phenomena~\cite{ciuti} that have attracted more and more attention
to the field of polaritonics in inorganic semiconductor
microcavities since the pioneering observation of the strong
coupling regime.~\cite{weis}

The weak binding energy and oscillator strength of Wannier-Mott
excitons characteristic of inorganic semiconductors are limitations that can be
overcome employing organic semiconductors having strongly bound
Frenkel excitons with a large oscillator
strength.~\cite{organic_exc} The strong coupling regime in an
organic based microcavity was first observed at room temperature using a
porphirine molecule (4TBPPZn) dispersed in a polysterene film as
optically resonant material at room temperature,~\cite{lidzey} and
later in a variety of organic materials~\cite{oulton,
organic_exc} including polyacene molecular
crystals.~\cite{Holmes,prlforrest2,rayleigh} 
The latter are also
characterized by the presence of well developed vibronic replicas
that participate in polariton formation and 
affect their luminescence.~\cite{LeoLuca2} 
In contrast to the case of inorganic microcavities, manifestations of
bosonic stimulation using organic cavity polaritons have been quite
elusive. 
Recently, however, several non-linear phenomena were reported:
room temperature polariton lasing in
an anthracene single crystal microcavity,~\cite{Lasing} 
indirect pumping of J-aggregate lasing microcavities,~\cite{JaggregatesLasing} 
and non-linear emission in polymer-based microcavities.~\cite{Mahrt}
In anthracene, the observation of a threshold for
nonlinear emission was accompanied by a significant line
narrowing and by a collapse of the emission lifetime. 
In that case, a comparison with the best-case estimate of the threshold for
conventional lasing inferred from amplified stimulated emission
measurements shows that the lasing threshold
observed in the strongly coupled microcavity is slightly lower than that anticipated for a conventional laser.~\cite{Lasing} 
The temperature dependence of
the polariton lasing threshold has also been investigated and shows
an order of magnitude decrease from room temperature to low
temperatures.~\cite{LasTemp} These experiments demonstrate the
high excitation density regime of polariton bosonic stimulation,
which could pave the way  to the observation in organic based
microcavities of other phenomena related to polariton fluidics
where weak polariton-polariton interactions may also
manifest.~\cite{ciuti}

In the present work, we develop a semiclassical kinetic model to
describe the onset of the non-linear threshold for polariton
lasing in anthracene-based microcavities. We show, in particular,
that the mechanism providing the bosonic final-state stimulated
formation of the ensemble of lower cavity polaritons is the
vibrationally assisted radiative decay of incoherent excitons, previously
populated by non-resonant pumping. In Section~\ref{sec:me}, we set
up a minimal master equation to describe the polariton population
dynamics, we make a realistic choice of material parameters  and
we fit the experimental data on the pump dependence of the
polariton emission, pointing out the relevance of bimolecular
quenching processes. In Section~\ref{sec:scattering}, we calculate
microscopically the efficiency of the relevant scattering process
justifying the value obtained from the fit. In
Section~\ref{sec:temperature}, we consider within our model the
dependence of the polariton lasing threshold on temperature.
Finally, in Section~\ref{sec:conclusion}, we present our
conclusions.

\section{Two-Level Model}\label{sec:me}

We model the dynamics of the lasing process using a minimal rate-equation approach.
In this section, 
we estimate the typical time-scale of the mechanism which selectively transfers excitations from the reservoir to the bottom of the polariton branch, without any assumptions regarding its microscopic nature. 

\subsection{The Master Equation}

The anthracene crystal has two molecules per unit cell and strongly anisotropic optical properties.~\cite{Pope, prlforrest2}
Excitations in this material are well-described within the Frenkel-exciton framework, which is based on the intramolecular promotion of an electron from the highest occupied molecular orbital to the lowest unoccupied one. Because of molecular dipole-dipole interaction, the excitation can propagate, resulting in
two orthogonal transition dipole moments, $\vec \mu_{a,b}$, directed along the in-plane $\mathbf a$ and $\mathbf b$ axes.
When a thin anthracene crystal is placed between two mirrors, light couples to both $a$- and $b$-polarized excitons and creates two orthogonally-polarized lower polariton branches.
Measurements are usually reported for light polarized along $\mathbf a$ and $\mathbf b$:~\cite{Lasing, prlforrest2} in these cases the $p$ and $s$ in-cavity light polarizations separately couple to the dipole moments $\vec \mu_{a,b}$ and no mixing effect is present.

We focus only on $b$-polarized excitons,~\cite{Pope} i.e. those with largest oscillator strength, for which lasing has been reported~\cite{Lasing} and neglect other polaritonic and excitonic states.
The initial relaxation of the pump excitations is also neglected,
and the presence of an effective excitonic reservoir at a fixed energy independent on the cavity properties is considered.~\cite{Michetti1}
We note that the experimental photoluminescence (PL) from anthracene microcavities shows always a clear maximum at energy $ \sim 2.94$ eV regardless of the cavity thickness,~\cite{Lasing, Thesis} and indeed lasing has been achieved in a cavity where the minimum of the LP is exactly at $2.94$ eV.
This is a signature that the microscopic dynamics resulting in the lasing phenomenon is that of a two-level process rather than that of the well-known polariton bottleneck. 
We thus develop a two-level master equation for $\nu_e(t)$ and $\nu_p (t)$, the surface density of reservoir excitons and of lasing polaritons located near $\mathbf k = 0$, respectively. 

\begin{figure}[t]
 \includegraphics[width=\columnwidth]{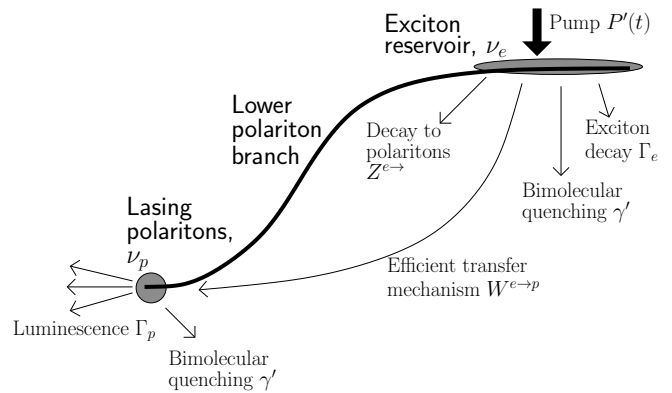}
 \caption{Sketch of the LP branch and of the physical processes and scattering mechanisms included in master equation~(1).}
 \label{fig:sketch}
\end{figure}

We denote with $ \mathcal A_0$ the subregion of the Brillouin zone located around $\mathbf k =0$ which is occupied by the lasing polaritons. Because states at the bottom of the LP branch do not have a well-defined wavevector $k$, we consider all of the localized wavepackets with energy $\sim E_{\rm LP}(\mathbf k=0)$ as equally contributing to the lasing process.
$N_{\rm pol}$ is the number of such polaritonic states, while $N_{\rm exc }$
is the number of excitonic states.
The polariton and exciton decay rates are $\Gamma_p = |c^{(e)}_p|^2/ \tau_e + |c^{(p)}_p|^2/ \tau_{p}$ and $\Gamma_e= 1/\tau_e$, respectively, where $\tau_{p }$ $(\tau_{e })$ is the bare photon (exciton) lifetime and $c_{p}^{(p)}$ $\left(c_{p}^{(e)} \right)$ is the photonic (excitonic) Hopfield coefficient for the lasing polaritons.

The parameter $Z^{e \rightarrow}$ is the decay rate via other channels, such as phonons, lower polaritons outside the $\mathcal A_0$ region and leaky modes, whereas bimolecular quenching processes are treated separately, with a rate $\gamma'$. 
A standard pump term proportional to $P'(t)$ is included; in order to take into account possible saturation effects the term $(1- \nu_e /\bar{\nu}_e)$ has been considered ($\bar \nu_e = N_{\rm exc}/A$ is the surface-density of excitonic states and $A$ is the area of the sample).

The rate of resonant excitation transfer from the reservoir to the lasing polaritons is $W^{e \rightarrow p}$.  We retain the bosonic enhancement term $(1 + \nu_p / \bar{\nu}_p)$ responsible for lasing effects, where $\bar \nu_p = N_{\rm pol}/A$ is the surface-density of polaritonic states.

The master equation for $\nu_e(t)$ and $\nu_p(t)$, whose physics is sketched in Fig.~\ref{fig:sketch}, reads:
\begin{subequations}
\begin{eqnarray}
\label{eq:experimentalS1}
\dot{\nu}_e &=& - \Gamma_e \nu_e - W^{e \rightarrow p}\nu_e  \left( 1+ \frac{\nu_p}{\bar{\nu}_p}  \right) - Z^{e \rightarrow} \nu_e + \nonumber\\
&&- \gamma'   \left( \nu_e +   |c^{(e)}_p|^2 \nu_p \right)\nu_e+ \left(1- \frac{ \nu_e}{\bar{\nu}_e} \right) P'(t) \\ 
\dot{\nu}_p &=& - \Gamma_p \nu_p +  W^{e \rightarrow p} \nu_e \left(1+ \frac{\nu_p}{\bar{\nu}_p} \right) + \nonumber \\
&&- \gamma'  \left(  \nu_e +   |c^{(e)}_p|^2 \nu_p \right) |c^{(e)}_p|^2  \nu_p \qquad 
\label{eq:experimentalS2}
\end{eqnarray}
\end{subequations}
The full derivation is given in appendix~\ref{sec:master}.
Note that the resulting equations are completely analogous to those describing conventional lasing,~\cite{Lasers} with the important difference that the lasing state is a polariton and thus retains an excitonic component.

\begin{table}

Simulation Parameters

\smallskip

\begin{tabular}{c|c}
\toprule
$\rho_0 = 4.2 \times 10^{21} $ cm$^{-3}$ & $L_z = 120$ nm\\
\colrule
$\bar \nu_e = 5.4 \times 10^{-16}$ cm$^{-2}$&$q_0 = 2.2 \times 10^4$ cm$^{-1}$ \\
\colrule
$\tau_p = 85$ fs $\sim 1$ ps & $\tau_e = 2$ ns \\
\colrule
$c_p^{(p)}= 0.92$ & $c_p^{(e)}= 0.39$ \\
\botrule
\end{tabular}

\smallskip \smallskip \smallskip

Fit Parameters

\smallskip

\begin{tabular}{c||c|c}
\toprule
$\tau_p$ & $W^{e \to p}$ & $\gamma'$ \\
\colrule
$85$ fs & $4 \times 10^5$ s$^{-1}$ & $1.5 \times 10^{-5}$ cm$^2 s^{-1}$ \\
$1$ ps & $3.5 \times 10^4$ s$^{-1}$ & $1.6 \times 10^{-5}$ cm$^2 s^{-1}$ \\
\botrule
\end{tabular}
\caption{Parameters for the numerical simulations. 
(top) List of the most important simulation parameters used in the numerical simulations. (bottom) Results of the fit reported in Fig.~\ref{fig:Simple2}}
\label{tab:PARAMETERS}
\end{table}

\subsection{Parameters}

We relate Eq.~(1) to the experimental system in Ref.~[\onlinecite{Lasing}] using the following parameters. 

\paragraph{Anthracene Crystal.}
The experimental microcavity embeds a crystal of anthracene with thickness $L_z = 120$ nm; the molecular density is $\rho_0 =4.2 \times 10^{21}$ cm$^{-3}$: we ignore the monoclinic structure of the unit cell and instead estimate its linear size as  $a = (\rho_0/2)^{-1/3} = 7.8 \times 10^{-8}$ cm, including the presence of two molecules per unit cell. 
The number of layers is estimated as $N =L_z/a \approx 153$. The absorption maximum of the anthracene crystal is at energy $E_0 = 3.17$ eV. The exciton measured lifetime  is of the order of $\tau_e \sim 1 -3$ ns and in the next simulations we take the intermediate value $\tau_e = 2$ ns. 
The contribution of $Z^{e \rightarrow}$ is neglected because it can be included into $\tau_{e}$ without any substantial difference as long as $\tau_{e} < 1/Z^{e \rightarrow}$, which can be safely assumed.

\paragraph{Microcavity and Polaritons.}
If we assume homogeneous broadening, the cavity lifetime can be estimated from the polariton linewidth at $\mathbf k=0$, where it is mostly photon-like. Using this approach, we obtain a lower bound $\tau_{p}=85$ fs. 
An exact calculation assuming perfect interfaces for the mirrors results in an upper bound $\tau_p = 1$ ps. 
We will estimate $W^{e \rightarrow p}$ corresponding to both extrema.
The Hopfield coefficients of the LP branch are~\cite{prlforrest2}: $c_{p}^{(p)} = 0.92$ and $c_{p}^{(e)} = 0.39$.

For small $|\mathbf k|$, the $\mathcal A_0$ region has cylindrical symmetry,~\cite{maxwellanisotropo,zoubi,rayleigh}.  
Its radius, $q_0$, can be estimated using 
$E_{\rm LP}(q_0)- E_{\rm LP}(k=0) = \Gamma_{0} / 2$, 
where $\Gamma_0 = 15$ meV is the linewidth of polaritons at $\mathbf k=0$ below threshold;~\cite{Lasing}
we obtain $q_0 = 2.2 \times 10^4$ cm$^{-1}$. 

\begin{figure}[t]
\includegraphics[width=0.8\columnwidth]{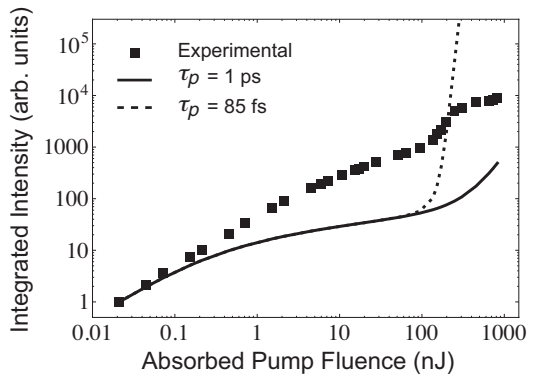}
\caption{Time-integrated surface density of polaritons $\int \nu_p (\tau) d\tau$ calculated from solution of Eq.~(1) (lines) and from experimental data (squares). The bimolecular quenching rate is taken from the measured 3D bulk value: $\gamma' = \gamma_{\rm 3D}/L_z$.  The calculation parameters are: (dashed line)  $\tau_{p} = 85$ fs, $W^{e \rightarrow p} = 7 \times 10^5$ s$^{-1}$; (solid line) $\tau_p = 1$ ps, $W^{e \rightarrow p} = 7 \times 10^4$ s$^{-1}$. Because the experimental data is in arbitrary units, here and in the following fits the experimental data is normalized so that the first experimental and theoretical points coincide.}
\label{fig:Simple}
\end{figure}

\paragraph{Pump.} 
The pump density is: $$P'(t) = P_0' \exp \left[ - \frac{t^2}{2 \sigma^2} \right]; \quad \sigma = \frac{150}{2 \sqrt{2 \ln 2}} \, \text{fs} \approx 64 \, \text{fs},$$ with $P'_0 = P_0/(\pi r_0^2 \hbar \omega_{\rm pump})$ where $r_0 = 110 $ $\mu$m is the radius of the pump spot and $\hbar \omega_{\rm pump} = 3.45$ eV is the energy of the pump photons. 
Because $E_{\mathrm{tot}} = \int P(t) dt = \sqrt{2 \pi} P_0 \sigma$  and because $E_{\rm tot}$, the total absorbed energy, and $\sigma$ are experimentally known, $P_0$ is also known. 

\paragraph{Bimolecular Quenching Rate.}
To the best of our knowledge there are no measurements of the bimolecular quenching rate, $\gamma'$, for two-dimensional anthracene crystals.
According to the standard theory for bimolecular quenching~\cite{Pope}, $\gamma_{\rm 3D} = 8 \pi R D$, where $R$ is the F\"orster radius of the exciton and sets the volume around the exciton in which annihilation happens,
while $D$ is the diffusion coefficient of excitons. 
Measurements for three-dimensional anthracene crystals have yielded values of~\cite{Pope} $\gamma_{\rm 3D} = 10^{-8}$ cm$^3$ s$^{-1}$ and~\cite{Pope, Diffusion} 
$D \sim 1 - 10 \times 10^{-3}$ cm$^2$ s$^{-1}$. 
The corresponding diffusion length $\ell = (\tau_e D)^{1/2} \sim 1 - 3 \times 10^{-6}$ cm is smaller than $L_z = 1.2 \times 10^{-5}$ cm and suggests that excitons can be treated as diffusing in a three-dimensional environment.
As a result, we initially fix $\gamma' = \gamma_{\rm 3D} /L_z = 7 \times 10^{-4}$ cm$^{2}$ s$^{-1}$.

Parameters used in the numerical simulations are briefly summarized in Table~\ref{tab:PARAMETERS}.

\subsection{Results}

\begin{figure}[t]
\includegraphics[width=0.8\columnwidth]{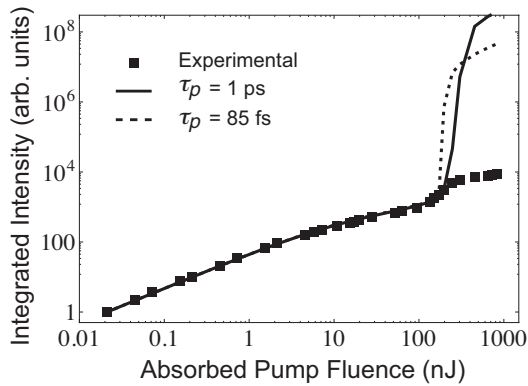}
\caption{Time-integrated surface density of polaritons $\int \nu_p (\tau) d\tau$ calculated from solution of Eq.~(1) (lines) and from experimental data (squares). The bimolecular quenching parameter $\gamma'$ is used to fit the below-threshold behavior of the experimental data.
The calculation parameters are: (dashed line) $\tau_p = 85$ fs,  $\gamma' = 1.5 \times 10^{-5} $ cm$^2$ s$^{-1}$, $W^{e \rightarrow p} = 4 \times 10^5$ s$^{-1}$; (solid line) $\tau_p = 1$ ps, $\gamma' = 1.6 \times 10^{-5} $ cm$^2$ s$^{-1}$, $W^{e \rightarrow p} = 3.5 \times 10^4$ s$^{-1}$.  }
\label{fig:Simple2}
\end{figure}

Since all other parameters are known, we leave only $W^{e \rightarrow p}$ as a fit parameter. 
We numerically integrate equations~\eqref{eq:experimentalS1} and~\eqref{eq:experimentalS2} and once the complete time-dependent functions $\nu_{e,p}(t)$ are known we compute the integral 
$\int \nu_p(\tau) \, {\rm d}\tau$ 
and compare it with the experimental values.

In Fig.~\ref{fig:Simple} the fits obtained for the extreme values of $\tau_{p} = 85$ fs and $1$ ps are shown. 
The value of $W^{e \rightarrow p}$ has been fit to the experimentally observed threshold value. 
In both cases, $W^{e \rightarrow p}$ is of the order $10^5$ s$^{-1}$. 
The agreement with the experiment is poor and it is apparent that the chosen value of $\gamma'$ does not properly describe the transition between linear and sublinear region below threshold.
Note that the exciton lifetime $\tau_e \sim 2$ ns is shorter than the reported bulk value~\cite{Pope} $\tau_{e, \mathrm{bulk}} \sim 10$ ns; surface interactions or defects within the layers 
could explain this discrepancy. 
In this situation, the excitonic diffusion coefficient can be smaller, resulting in a reduced possibility for excitons to pairwise annihilate. 

\begin{figure}[t]
\includegraphics[width=0.9\columnwidth]{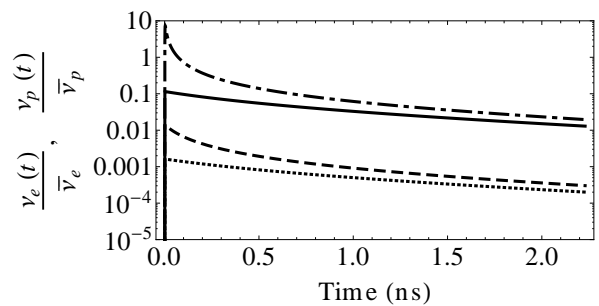}
\caption{Time-dependence of the normalized surface density of excitons $\nu_e(t) / \bar \nu_e$ and of polaritons $\nu_p (t) / \bar \nu_p$ below threshold ($E_{\rm tot} = 17 $ nJ, dotted line and solid line respectively) and at threshold ($E_{\rm tot} = 150 $ nJ, dashed line and dashed-dotted line) plotted for $\tau_p =85$ fs. Note that this time dependence is in good agreement with that reported in Ref.~[\onlinecite{Lasing}].
}
\label{fig:pop:time}
\end{figure}

Because a fit of $\gamma'$ which determines the onset of bimolecular quenching can be readily decoupled from that of $W^{e \rightarrow p}$,
both parameters are allowed to vary and the resulting fits are shown in Fig.~\ref{fig:Simple2}.
We obtain $\gamma' \approx 1.5 \times 10^{-5}$ cm$^2$ s$^{-1}$ 
independently of $\tau_p$, as expected.
Note that this value is two orders of magnitude smaller than 
$\gamma_{\rm 3D} /L_z$.
The resulting values for $W^{e \rightarrow p}$ are $ 4 \times 10^5$ s$^{-1}$ and $3.5 \times 10^4$ s$^{-1}$ for $\tau_p = 85$ fs and $1$ ps, respectively.
Even if the scattering process acts on a sensibly longer timescale compared to the exciton and polariton lifetimes, it can lead to observable effects in presence of high excitonic densities. We can roughly estimate the surface density of excitons at threshold via $\Gamma_p \nu_p =  W^{e \to p} \nu_e (1 + \nu_p / \bar \nu_p)$.
Assuming that at threshold $\nu_p = \bar \nu_p$, we obtain:
\begin{displaymath}
 \frac{\nu_{e, th}}{\bar \nu_e} = \frac{\bar \nu_{p}}{\bar \nu_e} 
\frac{\Gamma_p}{ 2 W^{e \to p}} \sim 0.01
\end{displaymath}
The density of excitations is thus extremely high, though not unrealistic.
Moreover, this is consistent with what shown in Fig.~\ref{fig:pop:time}, where at threshold the peak exciton value is of few percents.
Note that $\nu_{e, th} / \bar \nu_e$ does not depend on the value of $q_0$, because both $W^{e \to p}$ and $\bar \nu_p$ depend linearly on the size of the $\mathcal A_0$ region.

\begin{figure}[t]
\includegraphics[width=0.9\columnwidth]{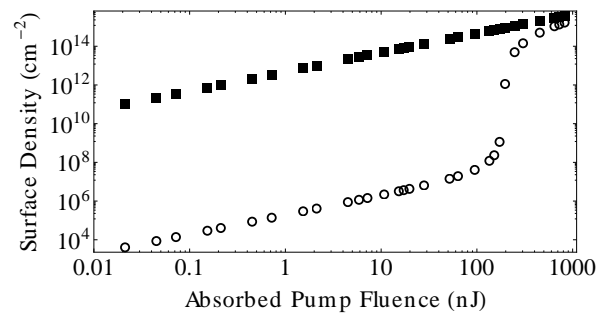}
\caption{Maximal population density of excitons $\max_t \nu_e(t)$ (squares) and of lasing photons $\max_t \nu_p(t)$ (circles).
See Fig.~\ref{fig:Simple2} for the parameters; $\tau_p =85$ fs.}
\label{fig:max:pop}
\end{figure}

Although the fit below threshold is excellent, the region above threshold is poorly described.
It can be seen in
Fig.~\ref{fig:pop:time} and Fig.~\ref{fig:max:pop}, which shows the time dependence and peak of the
normalized surface exciton and polariton densities, that at threshold the exciton density reaches a few percent of the total molecular density. Such high excitation densities may require a more refined description of the annihilation process. Indeed, our calculation above threshold seems to be in better agreement with recent low-temperature data, where the threshold occurs at lower excitation density.~\cite{LasTemp} Moreover, above threshold, when the polariton density becomes important, the details of the theoretical model used for the polariton-polariton bimolecular quenching become important.~\cite{bulovic, AgraBMQ}
Note that the mean-field polariton-polariton interaction~\cite{Zoubi2} has not been included as no blue shift has been resolved in the experiments, which feature a relatively broad linewidth.~\cite{Lasing,LasTemp}

In conclusion, using our simple two-level model we have extracted an estimate for the scattering process $W^{e \to p}$ relevant to polariton lasing in anthracene. Furthermore, we believe that the strongly reduced rate of bimolecular annihilation observed should motivate further experimental and theoretical studies of this process.

\section{The Scattering Mechanism}\label{sec:scattering}

In this section we focus on the microscopic origin of the excitation transfer of Sec.~\ref{sec:me}. In particular, we propose  as the relevant mechanism the radiative recombination of a molecular exciton assisted by the emission of a vibrational quantum of the electronic ground state.~\cite{LeoLuca2}
We show that the resulting scattering rate is in good agreement with that obtained in the previous Section.
Finally, we also consider an alternative and possibly coexisting model based on the non-radiative emission of an optical phonon.~\cite{Litinskaya}

\subsection{Radiative Transition}

\begin{figure}[t]
\includegraphics[width=\columnwidth]{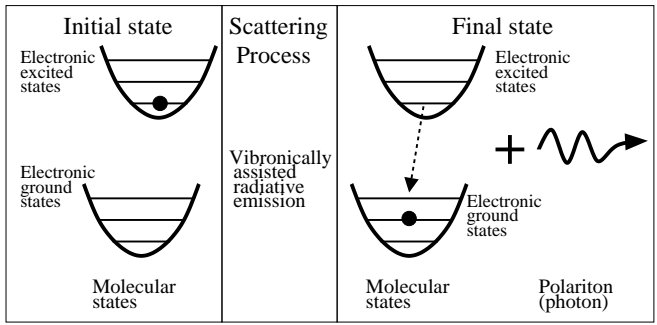}
\caption{Sketch of the radiative microscopic mechanism responsible for the efficient excitation transfer in the Franck-Condon approximation.}
\label{fig:micro}
\end{figure}

The absorption and PL spectra of anthracene show several vibronic resonances.~\cite{VibrQuantum} 
The resonances observed in absorption correspond to the molecular vibrations of the first electronically excited state, and those in PL to the vibrations of the electronic ground state.~\cite{Pope}
Strong light-matter coupling has only been demonstrated for the former,~\cite{Holmes} since the fraction of vibrationally excited ground-state molecules is negligible at room temperature.
However, as shown schematically in Fig.~\ref{fig:micro}, the transitions responsible for the vibronic structure in PL result in the scattering of excitons to lower energy polaritons, where the missing energy exactly corresponds to that of the vibrational quantum, $E_{01} \sim 173$ meV.~\cite{LeoLuca2}

In appendix~\ref{app:scattering}, we discuss the approximations needed to apply the known microscopic theory~\cite{zoubi, LeoLuca2} to the present system~\cite{prlforrest2}. For instance, the microscopic model considers a thin organic material comprising $N \sim O(1)$ layers and placed in the middle of the microcavity~\cite{zoubi,LeoLuca1} whereas the experimental sample embeds an organic material which fills the whole region between the two mirrors ($N \sim O(10^2)$) and has no planar translational invariance.~\cite{Lasing} Moreover, the theory assumes the presence of perfect mirrors, whereas in  experiment $\tau_p$ is always finite.

In equation~\eqref{eq:important} the scattering rate $W^{j \to k}$ from one molecular exciton (labelled by $j$) to a lasing polariton (labelled by $k$) is related to the parameters of a anthracene microcavity.
As discussed in appendix~\ref{sec:master}, the scattering rate appearing in the master equation is $W^{e \rightarrow p} = \sum_{ k \in \mathcal A_0}  W^{j \rightarrow  k}$. Working in the energy space and defining the spectral region of lasing polaritons $E \in [E^{\mathcal A_0}_{\rm inf}, E^{\mathcal A_0}_{\rm sup}]$ and the polariton density of states $D(E)$, we get:
\begin{equation}
W^{e \rightarrow p} =
 \int_{E^{\mathcal A_0}_{\rm inf}}^{E^{\mathcal A_0}_{\rm sup}}
\frac{ V_{1 \pm}^2 }{\hbar}
\, \frac{\pi^2 S  |c^{(p)}_{p} |^2 }{2 MN} \,  f (E_0 - E - E_{01}) D(E) {\rm d}E 
\label{eq:scattering0}
 \end{equation}
$M$ the number of unit cells in the two-dimensional quantization surface and $V_1$ is the fit light-matter coupling. $S$ is the Huang-Rhys parameter, which is approximately $\sim 1$.~\cite{Spano2} 
See appendix~\ref{app:scattering} for more details.
The 2D density of state is
$$D(E) = \frac{m M a^2 }{2 \pi \hbar^2 } \; 
\theta\left[ \, E-E_{\rm LP}(\mathbf k=0) \, \right],$$ 
where the effective mass $m$ can be obtained from the fits of the dispersion relations: 
$m \sim 1.7 \times 10^{-5} m_e$; moreover $\bar{\nu}_e = 2 N/a^2 $. 
The normalized linewidth of (0-1) photoluminescence $f(E)$ is a Lorentzian centered in zero with FWHM $\Gamma = 100$ meV; 
we also assume $E_{\rm inf}^{\mathcal A_0}=E_0 - E_{01} = E_{\rm LP}(\mathbf k = 0)$
whereas $E_{\rm sup}^{\mathcal A_0 } = E_{\rm inf}^{\mathcal A_0}+7.5$ meV (see Sec.~\ref{sec:me}).
We obtain:
\begin{equation}
W^{e \rightarrow p} =
 \frac{\pi S |c^{(p)}_{p} |^2 }{2 }  \frac{m V_{1 }^2}{\bar \nu_e \hbar^3 } \int_{E_{\rm inf}^{\mathcal A_0}}^{E_{\rm sup}^{\mathcal A_0}} f(E_{\rm inf}^{\mathcal A_0}-E) {\rm d}E.
\label{eq:scattering}
\end{equation}
The rate before the integral is equal to $\approx 1.0 \times 10^{7}$~s$^{-1}$, while the contribution from the integral, which comes from the lineshape, is $\approx 0.047$.
Thus, the theoretical microscopic mechanism is 
$W^{e \rightarrow p} \approx 5 \times 10^5$ s$^{-1}$. 

Because the theoretical model neglects effects which can possibly lower the efficiency of the resonant scattering, we consider our estimate to be in good agreement with the values estimated from data in Sec.~\ref{sec:me}. 

\subsection{Non-radiative transition}

We now consider an alternative and possibly coexisting relaxation channel, 
which is non-radiative.~\cite{Litinskaya}
An exciton is scattered from the reservoir to one polariton state by the emission of a molecular vibration of the electronic excited state.
This is due to the intramolecular exciton-phonon 
coupling~\cite{Soos, organic_exc} which has been demonstrated to play a key role in the modeling of the PL of J-aggregates microcavities.~\cite{PhononMichettiTH, PhononMichettiEXP}

Note that in this case the considered phonon belongs to the electronic excited state, whereas in the radiative case it was related to the electronic ground state. Moreover, the resulting scattering element $W^{e \rightarrow k}$ includes the excitonic content of the outcoming polariton, whereas Eq.~\eqref{eq:scattering} is weighted by the photonic Hopfield coefficient.

The scattering rate from one molecular exciton (labelled by $j$) to one lasing polariton (labelled by $k$) is given by:~\cite{Litinskaya}
\begin{equation}
W^{j \to k} =
\frac{2 \pi}{\hbar} g^2 E^2_{11} \frac{| c_{p}^{(e)} |^2}{2NM}
\delta (E_0 - E_{\rm LP}(\mathbf k)-E_{11})
\label{eq:NonRadiative}
\end{equation}
where $g=\sqrt S \sim 1$ is the strength of the exciton-phonon coupling,~\cite{Spano2} 
$E_{11}$ is the energy quantum of a vibration of the excited state. 
Even if the Franck-Condon model which we are using prescribes $E_{11}=E_{01}$, this is not necessarily true in general. 
The factor $| c_{p}^{(e)} |^2/(2NM)$ is the Hopfield coefficient for the exciton of the  molecule $j$ relative to the polariton $k$.~\cite{Litinskaya} 
Because 
$c_p^{(e)} = c_k^{(e)} = \sum_j c_k^{(j)}$ $\forall k$,
we are assuming that the exciton is equally distributed among all the molecules.
This is consistent with the assumptions used in the derivation of the master equation (see appendix~\ref{sec:master}).

The comparison of Eq.~\eqref{eq:NonRadiative} with Eq.~\eqref{eq:B5} for the radiative case shows that the two processes have a similar efficiency. Indeed, using Eq.~\eqref{eq:Vmicroscopic}:
\begin{equation}
 \dfrac{W^{j \to k}_{\rm RAD}}{W^{j \to k}_{\rm NON-RAD}} = 
 \dfrac{\pi V_1^2}{2 E_{11}^2} \cdot
 \dfrac{|c_p^{(p)}|^2}{|c_p^{(e)}|^2};
\end{equation}
because both $V_1$ and $E_{11}$ are of the same order of magnitude, $100$ meV, the efficiency ratio mainly depends on the Hopfield coefficients of the bottom polaritons.
Thus, as in our case $|c_p^{(p)}|^2/|c_p^{(e)}|^2 \approx 5$, we expect the radiative mechanism to be the main origin of the excitation transfer which results in lasing, even if to understand the importance of the non-radiative transfer a more detailed analysis is necessary.

In conclusion, we have studied two physical mechanisms which can possibly 
induce the excitation transfer studied in Sec.~\ref{sec:me}.
Using simple models, we have obtained estimates which are in good agreement with those from the data.
The photonic and excitonic components of the bottom polaritons are crucial for determining the importance of the two mechanisms. We thus expect that in materials requiring different cavity detunings to match the condition 
$E_0 - E_{01} = E_{\rm LP} (\mathbf k =0)$, 
the relevance of the two processes could be reversed.
An experimental analysis exploring  several organic crystals would thus be of the greatest interest.

\section{Temperature}\label{sec:temperature}

Reported data for anthracene microcavities show a reduction of the lasing threshold of slightly less than an order of magnitude once temperature is lowered from $300$~K to $12$~K.~\cite{LasTemp}
In this section we discuss temperature effects within the framework of the developed model, and the related consequences on the lasing properties.

\begin{figure}[t]
\includegraphics[width=\columnwidth]{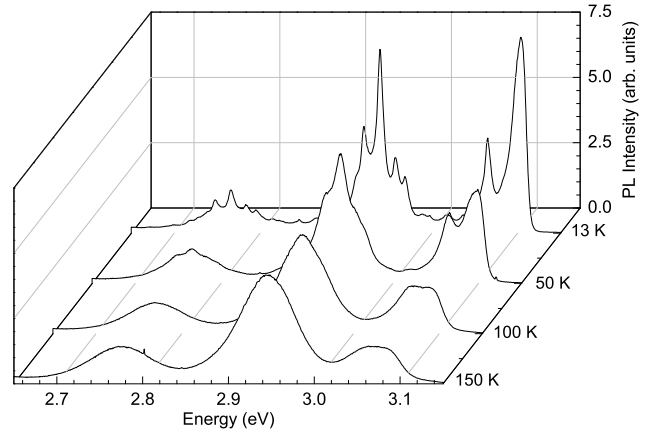}
\caption{Photoluminescence spectra of anthracene layers for temperatures between $13$ K and $150$ K.}
\label{fig:spectra}
\end{figure}

Experimental studies on the PL from bulk anthracene crystals have shown a strong temperature dependence characterized by considerable spectral narrowing.~\cite{Lyons} 
The temperature dependence obtained using thin crystals grown from solution is shown in Fig.~\ref{fig:spectra}. Here, the crystals were grown on silicon substrates to ensure good thermal contact to the cryostat cold finger and were excited using 1 ns-long pulses at $\lambda$=337 nm.
A composite vibronic structure emerges, which can be understood in terms of a high-energy phonon (considered in this work) and of a low-energy phonon, which is not resolved at room temperature because of thermal broadening.
Such a system requires the use of two-phonon states in order to exactly reproduce the spectra;~\cite{SilvestriSpano} however, we ignore this complication
because we are only interested in the phenomenological properties of the line which is responsible for lasing.

The scattering rate $W^{e \rightarrow p}$ in Eq.~\eqref{eq:scattering} depends on temperature via $f(E)$. On the one hand, at low temperature the Lorentzian is narrower, and thus a smaller fraction of the oscillator strength is dispersed into non-lasing modes. On the other hand, only a fraction of the oscillator strength of the (0-1) transition contributes to lasing, because the other lines are far detuned. Additionally, both the quantum yield, estimated at room temperature to be $0.5$, and the exciton lifetime $\tau_e$ are expected to increase at low temperature. 

\begin{figure}[t]
\begin{center}
\includegraphics[width=0.75\columnwidth]{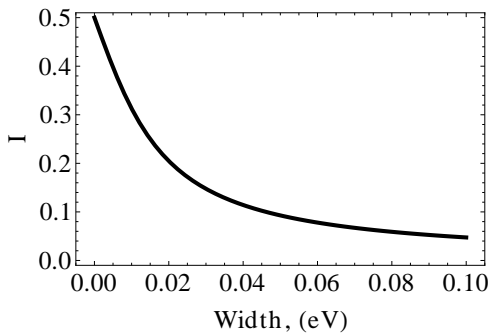} 
\caption{\label{fig:widthdep} Dependence of the integral~\eqref{eq:theintegral} appearing in~\eqref{eq:scattering} on the width of $f(E)$, which is a Lorentzian function.}
\end{center}
\end{figure}

In Fig.~\ref{fig:widthdep} we compute the dependence of the integral appearing in~\eqref{eq:scattering}:
\begin{equation}
I \doteqdot \int_{E^{\mathcal A_0}_{\rm inf}}^{E^{\mathcal A_0}_{\rm sup}} f(E_{\rm inf}^{\mathcal A_0}-E) \; {\rm d}E
\label{eq:theintegral}
\end{equation} on the width of the Lorentzian function $f(E)$ which represents the normalized spectrum of the (0-1) PL emission. Whereas at room temperature the FWHM is $\approx 0.1$ eV, at $12$ K it is $\sim 0.01 - 0.02$ eV, and thus $W^{e \rightarrow p}$ increases of at least a factor of $5$. 

Roughly speaking, the observed thermal reduction of the threshold is of less than one order of magnitude,~\cite{LasTemp} and thus similar to the numbers of our estimates. This points out a possible connection between the temperature dependence of the laser threshold and of the PL of anthracene crystals. 
A more systematic analysis, both theoretical and experimental, goes beyond the scope of this work, and will be the focus of future investigations.
As long as the thermal linewidth narrowing is considered, we observe that
when the radiative transition is not perfectly resonant with the lasing polaritons
it could even result in the opposite effect.

For the sake of completeness, in appendix~\ref{app:vibro:thermal} we include the thermal population of the vibrations of the molecular ground state in the master equation~\eqref{eq:experimentalS1} and demonstrate that it can be safely neglected.

\section{Conclusions}\label{sec:conclusion}

In this work, a minimal model to describe the polariton lasing
observed in crystalline anthracene microcavities has been
developed. Only the essential features of the physical processes
involved have been included: the incoherently pumped exciton
reservoir, the vibronically assisted radiative scattering from the
reservoir to the bottom of the lower polariton branch, the onset
of bosonic stimulation and the build-up of the polariton
population with increasing pump intensity, the polariton losses
through the mirrors and bimolecular quenching processes. All the
relevant material parameters, except from the bimolecular
quenching rate, have been determined independently from the
experimental data on the pump dependence of the polariton
emission~\cite{Lasing}. In particular, the efficiency of the
scattering mechanism here considered - which takes into account
the prominent role of vibronic replicas in the photophysics of
anthracene microcavities~\cite{LeoLuca2} - has been calculated
microscopically. The numerical simulations obtained are in good
agreement with the data and describe well the onset of the
non-linear threshold for polariton lasing. A possible reason for
the observed temperature dependence of the
threshold~\cite{LasTemp} has also been discussed. The present
model could be extended to include further ingredients, in
particular polariton-polariton scattering, and be applied to other
microcavity systems exhibiting pronounced vibronic replicas.

\begin{acknowledgements}

We thank V. M. Agranovich, D. Basko, S. Forrest, L. Silvestri, M. Slootsky 
for fruitful discussions and R. Fazio for encouragement.
Financial support from the European FP7 ICARUS program (grant
agreement N. 237900) is gratefully acknowledged.
LM is funded by Regione Toscana POR FSE 2007-2013.
SKC and PM respectively acknowledge the Imperial College JRF scheme 
and the Deutsche Forschungsgemeinschaft for financial support.

\end{acknowledgements}

\appendix

\section{Derivation of the Master Equations} \label{sec:master}

We present the derivation of the master-equation~(1).
We focus on the exciton reservoir (excitons are labeled by $j$) and on the polaritons in the $\mathcal A_0$ region, which are resonantly  populated by the reservoir (labeled by $k$). 
The dynamics of the system is described by $N_{\rm exc} + N_{\rm pol}$ coupled differential equations:
\begin{subequations}
\begin{align}
\label{eq:textbook1}
\dot n_j =& - \Gamma_j n_j - \sum_{k \in \mathcal A_0} W^{j \rightarrow k} n_j (1+ n_k) - \sum_{k \notin \mathcal A_0} W^{j \rightarrow k} n_j + \nonumber \\
&- \gamma  \left( \sum_{j'} n_{j'} + \sum_{k'} |c_{k'}^{(e)}|^2 n_{k'}\right) n_j + (1- n_j) P(t) ;\\
\dot n_k =& - \Gamma_k n_k + \sum_j W^{j\rightarrow k} n_j (1+n_k) \nonumber \\
&- \gamma  \left(  \sum_{j'} n_{j'} +  \sum_{k'} |c_{k'}^{(e)}|^2 n_{k'}\right)
|c_{k}^{(e)}|^2 n_k .
\label{eq:textbook2}
\end{align}
\end{subequations}
The term $ \sum_{k \notin \mathcal A_0} W^{j \rightarrow k} n_j$ describes excitons scattered to other polariton states via other decay mechanism, as for example, lattice phonons and luminescence. We don't include a similar term $\sum_{k' \notin \mathcal A_0} W^{k' \rightarrow k} n_{k'} (1+n_k)$ in equation~\eqref{eq:textbook2} because negligible compared to the efficient direct scattering from the reservoir.
The probability of annihilating an exciton (or polariton) because of bimolecular quenching is proportional to the total number of excitons $ \sum_{j'} n_{j'} + \sum_{k' \in \mathcal A_0} |c_{k'}^{(e)}|^2 n_{k'}$  (we neglect the minor contribution of polaritons $k' \notin \mathcal A_0$).  

In order to derive
the master equation for the surface density of excitations $\nu_e(t) = \sum_j n_j(t) /A$ and $\nu_p(t) = \sum_k n_k(t) /A$
we have to make the following assumptions.
We take $W^{j \rightarrow k}$ to be independent from $j$ and $k\in \mathcal A_0$, renamed $W$; the same holds for $\Gamma_j$, substituted by $\Gamma_e$, for $\Gamma_k$, renamed $\Gamma_p$, and for $c_{k'}^{(e)}$, renamed $c_{p}^{(e)}$. We introduce the quantities $W^{e \rightarrow p} = \sum_{k \in \mathcal A_0} W$ and $Z^{e \rightarrow} = \sum_{k \notin \mathcal A_0}W^{j \rightarrow k} $.
Finally, $n_j$ and $n_k$ are not expected to have a significant dependence on $j$ and $k$.
Clearly, this approach is more justified the more the $\mathcal A_0$ region is small.
We sum the equations~\eqref{eq:textbook1} and~\eqref{eq:textbook2}:
\begin{subequations}
\begin{align}
\sum_j \dot n_j =& - \sum_j \Gamma_j n_j - \sum_j \sum_{k \in \mathcal A_0} W^{j \rightarrow k} n_j (1+ n_k) + \nonumber \\
& - \sum_j \sum_{k \notin \mathcal A3_0} W^{j \rightarrow k} n_j + \nonumber \\
& 
- \gamma  \left( \sum_{j'} n_{j'} + \sum_{k'} |c_{k'}^{(e)}|^2 n_{k'}\right)\sum_j n_j
+ \nonumber \\
&+ \sum_j (1- n_j) P(t) \qquad \\
\sum_k \dot n_k =& - \sum_k \Gamma_k n_k + \sum_k \sum_j W^{j\rightarrow k} n_j (1+n_k) + \nonumber \\
& - \gamma  \left( \sum_{j'} n_{j'} + \sum_{k'} |c_{k'}^{(e)}|^2 n_{k'}\right) \sum_k|c_{k'}^{(e)}|^2 n_k
\end{align}
\end{subequations}
Using the listed assumptions, we obtain:
\begin{subequations}
\begin{align}
\dot{\nu}_e =& - \Gamma_e \nu_e - W^{e \rightarrow p} \nu_e \left( 1+\frac{\nu_p}{\bar{\nu}_p} \right)  - Z^{e \rightarrow} \nu_e + \nonumber\\
&- \gamma'   \left( \nu_e + |c^{(e)}_p|^2 \nu_p \right)\nu_e+ \left(1- \frac{ \nu_e}{\bar{\nu}_e} \right) P'(t) \\ 
\dot{\nu}_p =& - \Gamma_p \nu_p +  W^{e \rightarrow p} \nu_e \left(1+ \frac{\nu_p}{\bar{\nu}_p} \right) +\nonumber \\
& - \gamma'  \left( \nu_e + |c^{(e)}_p|^2 \nu_p \right) |c^{(e)}_p|^2  \nu_p 
\label{eq:meq:2}
\end{align}
\end{subequations}
which is written in terms of the
surface density of excitonic states $\bar{\nu}_e = N_{\rm exc}/A = L_z \rho$ and of polaritonic states $\bar{\nu}_p = N_{\rm pol}/A$,
of the pump rate density $ P'(t) = \bar{\nu}_e P(t)$ and of the quenching parameter $\gamma' = \gamma A$.

\section{Scattering Rate Due to Radiative Transition} \label{app:scattering}

We compute the scattering rate of a molecular exciton to a lasing polariton state via radiative emission assisted by the emission of a vibration 
(see Sec.~\ref{sec:scattering}). 

Linear optical properties of strongly-coupled microcavities can be quantitatively described with a simple model for the light-matter interaction which conserves the in-plane momentum:
\begin{equation}
 H_{\mathbf k} = \left(
 \begin{array}{cccc}
\hbar \omega_{\mathbf k} & V_1 & V_2 & V_3 \\
V_1^* & E_{10} & 0 & 0 \\
V_2^* & 0 & E_{11} & 0 \\
V_3^* & 0 & 0 & E_{12}  
 \end{array}
 \right).
\label{eq:coupled}
\end{equation}
The energy of the cavity photon, 
$\hbar \omega_{\mathbf k} = (c/n_{\mathrm{eff}}) \sqrt{|\mathbf k|^2 + \pi^2/L^2}$, 
and the energy of the exciton accompained by $i$ vibronic replicas, $E_{1i}$, are measurable quantities. 
The couplings $V_i$ can be fit from the measured polaritonic dispersion relations,~\cite{prlforrest2} which are the eigenvalues of~\eqref{eq:coupled}.

We focus on the $b$ exciton and on light polarized along $\mathbf b$;
the microscopic expression of their coupling is:\cite{zoubi, LeoLuca1}
\begin{equation}
V^{\rm m}_{1} (\mathbf k) = \frac{ \mu e^{-S/2}  }{n_{\mathrm{eff}}} \sqrt{ \frac{8 \pi \hbar \omega_{\mathbf k} }{L a^2}} 
\sqrt{ \frac{2 (N+1)}{\pi} } \sqrt{1 - \frac{|\mathbf k|^2}{\frac{\pi^2}{L^2} + |\mathbf k|^2  }}
\end{equation}
where $S$ is the Huang-Rhys parameter, $\mu$ is the dipole moments of the $b$ Davidov branch, $n_{\mathrm{eff}}$ is the effective refractive index,
$\hbar \omega_{\mathbf k}$ is the photon energy, $L$ is the effective length of the cavity, $a$ is the spacing between molecules, $N$ is the number of monolayer comprising the organic material.
Neglecting the dependence on $\mathbf k$, we identify the fit parameter $V_1$ of equation~\eqref{eq:coupled} with the following microscopic expression:
\begin{equation}
V_1 = V^{\rm m}_{1 } (\mathbf k = 0) =  4  \mu  e^{-S/2} 
 \frac{  (\pi c \hbar)^{1/2}}{L \, a \, n_{\mathrm{eff}}^{3/2}} 
 \, (  N+1 )^{1/2} .
\label{eq:Vmicroscopic}
\end{equation}
As a simple consistency check of~\eqref{eq:Vmicroscopic}, we take $n_{\rm eff} = 1.74$ and $L = 120$ nm and $\mu \sim 1$ D: we obtain $V_1 \sim 74$ meV, whose order of magnitude is compatible with the fit value of $108$ meV.~\cite{prlforrest2}
Thus, even if the theoretical estimate is based on the assumption of a perfect cavity without losses, whereas the fit value refers to a realistic imperfect system, the error is under control.

Let's focus on the light-matter interaction responsible of the exciton radiative recombination assisted by the emission of one molecular vibration:
\begin{eqnarray}
\hat V_{\mathbf n} &=& - \mu \left( -\sqrt{S} e^{- \frac S2} \right) \hat v^{\dagger}_{\mathbf n} \hat B_{\mathbf n} 
\cdot \nonumber \\
& \cdot &
\left(
\sum_{\mathbf k}  \sqrt{\frac{4 \pi \hbar \omega_{\mathbf k}}{L a^2 M n^2_{\mathrm{eff}}}} \frac{\omega_{\bf k=0} }{ \omega_{\mathbf k}}
e^{- i \mathbf{k \cdot n_{||}}} \hat a^{\dagger}_{\mathbf k p} \right) + H.c. \qquad
\end{eqnarray}
where  $M$ is the number of unit cells included in the two-dimensional quantization area, $v_{\mathbf n}^{\dagger}$ is the operator creating a vibronic replica at the ground state of the molecule placed at $\mathbf n$, $B_{\mathbf n}$ is the operator destroying an electronic excitation, $\hat a_{\mathbf k p}^{\dagger}$ is the photon field operator with $p$ polarization.

We are interested in the scattering of the molecular exciton at $\mathbf n$ into the lasing polariton region $\mathcal A_0$. 
Using the Fermi Golden Rule, the scattering rate from one molecular exciton (labelled by $j$) and a lasing polariton (labelled by $k$) is (see also appendix~\ref{sec:master}):
\begin{equation}
W^{ j \rightarrow k} =
\frac{2 \pi}{\hbar} \frac{ \pi S}4 
\frac{ 16 \mu^2  e^{-S} \pi c \hbar }{L^2 a^2  n_{\mathrm{eff}}^3} 
\frac 1M
|c^{(p)}_{p} |^2   
\delta (E_0 - E_{\rm LP} (\mathbf k ) - E_{01})
\label{eq:B5}
\end{equation}
We are assuming that the scattering process only depends on the energy of the final state accordingly with the picture of bottom polaritons as states with a non-defined wavevector and with similar optical properties.~\cite{reviewlarocca, MichettiLarocca}

We can make the previous equation more realistic by substituting the delta function $\delta(E)$ with the normalized lineshape of the (0-1) photoluminescence, dubbed here $f(E)$. In this work, we consider a Lorentzian linewidth 
$f(E) = \Gamma/(2 \pi (E^2-(\Gamma/2)^2)  ) $.
Comparing this last expression to~\eqref{eq:Vmicroscopic} we get ($N+1 \approx N$):
\begin{equation}
W^{j \to k} \approx \frac{ V_{1 }^2 }{\hbar}
\, \frac{\pi^2 S  |c^{(p)}_{p} |^2 }{2} \frac 1 {M N}  \,  f (E_0 - E_{\rm LP} (\mathbf k) - E_{01})
\label{eq:important}
\end{equation}
This expression links the scattering rate assisted by the emission of one molecular vibration to known parameters.

\section{Thermal Population of Vibronic Replicas}\label{app:vibro:thermal}

\begin{figure}[t]
\includegraphics[width=0.8\columnwidth]{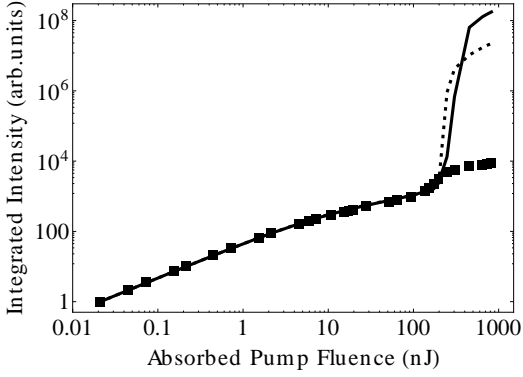}
\caption{Time-integrated surface density of polaritons $\int \nu_p (\tau) d\tau$ calculated from solution of Eq.~(C5) (lines) and from experimental data (squares). The calculation parameters are: (dashed line)  $\tau_{p} = 85$ fs; (solid line) $1$ ps. The fit parameters are as in Fig.~\ref{fig:Simple2}.}
\label{fig:App_S}
\end{figure}

Up to now the scattering of one polariton to the exciton reservoir assisted by the absorption of a replica of the ground state has been neglected. 
However, at room temperature, a fraction of the molecules quantified by the Bose-Einstein distribution is in a vibrationally excited state; 
taking $E_{01} \approx 173$ meV and room temperature ($k_B T \approx 25.6 $ meV) the 2D density of such molecules is:
\begin{equation}
 \bar \nu_e\times \frac 1{e^{E_{01}/k_B T} -1} \approx 5 \times 10^{16} \text{ cm}^{-2} \times  10^{-3} \approx 5 \times 10^{13} \text{ cm}^{-2}.
\end{equation}
Even if $10^{-3}$ is a small fraction in absolute terms, 
the density of phonon-excited molecules is comparable to the density of excitons of the previous simulations (see e.g. Fig.~\ref{fig:max:pop}).
Thus, polariton depletion because of back-scattering into the exciton reservoir can affect the gain of the lasing process. 

In order to study the effect of this process, we include
in the right-hand side of equation~\eqref{eq:textbook2} the term:
\begin{equation}
 - \sum_j W^{j \to k} m_j n_k,
\end{equation}
where the sum is over all the molecules and $m_j$ is the population of the phonon state of the $j-$th molecule. We do not consider $m_j$ as a dynamical variable but rather consider the thermal equilibrium population: 
$m_j \doteqdot (e^{ E_{01} / k_B T}-1)^{-1}$.
Consequently, Eq.~\eqref{eq:meq:2} includes the term:
\begin{equation}
 - \frac 1 A \sum_{k\in\mathcal A_0} \sum_j  W^{j \to k} m_j n_k =
 - W^{e \to p} \frac{\bar \nu _e}{e^{E_{01} / k_B T}-1}
 \frac{\nu_p(t)}{\bar \nu_p}.
\end{equation}
The depletion rate is estimated as:
\begin{equation}
 - W^{e \to p} \frac {\bar \nu_e} {e^{E_{01}/k_B T} -1}  \frac{\nu_p(t)}{\bar \nu_p} \approx - 10^{5} \times 10^{9} \times 10^{-3}  \text{ s}^{-1} \times \nu_p(t).
\end{equation}
We compare it to the polariton decay rate, $\Gamma_p > 10^{12}$ s$^{-1}$, and conclude that it is not the dominant polariton depletion mechanism.
This would be the case for microcavities with larger Q factors, which thus would benefit from lower temperatures freezing the main polariton decay channel.
The contribution of this process on the reservoir population is also negligible, because polaritonic states $N_{\rm pol}$ are a negligible fraction of the excitonic states $N_{\rm exc}$.

We now take into account the population of phonon-excited molecules $\nu_v(t) \doteqdot \sum_j m_j(t) / A$ dynamically.
We consider the following master equation (the derivation is a generalization of the previous discussion):
\begin{subequations}
\begin{eqnarray}
\label{eq:S1}
\dot{\nu}_e &=& - \left( \Gamma_e +Z^{e \rightarrow} \right)\nu_e - W^{e \rightarrow p}\nu_e  \left( 1+ \frac{\nu_p}{\bar{\nu}_p}  \right) + W^{e \to p}  \frac{\nu_p}{\bar \nu_p} \nu_v+ \nonumber\\
&&- \gamma'   \left( \nu_e +   |c^{(e)}_p|^2 \nu_p \right)\nu_e+ \left(1- \frac{ \nu_e}{\bar{\nu}_e} \right) P'(t) \\ 
\label{eq:S2}
\dot{\nu}_p &=& - \Gamma_p \nu_p +  W^{e \rightarrow p} \nu_e \left(1+ \frac{\nu_p}{\bar{\nu}_p} \right) - W^{e \to p}  \frac{\nu_p}{\bar \nu_p} \nu_v+ \nonumber \\
&&- \gamma'  \left(  \nu_e +   |c^{(e)}_p|^2 \nu_p \right) |c^{(e)}_p|^2  \nu_p \\
\dot{\nu}_v &=& - \Gamma_v \left(\nu_v - \frac{\bar{\nu}_e}{e^{E_{01}/kT}-1} \right)  
 - W^{e \to p}  \frac{\nu_p}{\bar \nu_p}  \nu_v  +\nonumber \\ 
&& + W^{e \rightarrow p} \nu_e \left(1+ \frac{\nu_p}{\bar{\nu}_p} \right)
\label{eq:S3}
\end{eqnarray}
\end{subequations}
$\Gamma_v$ models the relaxation to the vibrational ground state, and the presence of an equilibrium population is taken into account; we set $\Gamma_v = 10$ ps. 
The other two terms of equation~\eqref{eq:S3} are due to polariton back-scattering to the exciton reservoir and to the exciton radiative recombination respectively.

\begin{figure}[t]
\includegraphics[width=0.9\columnwidth]{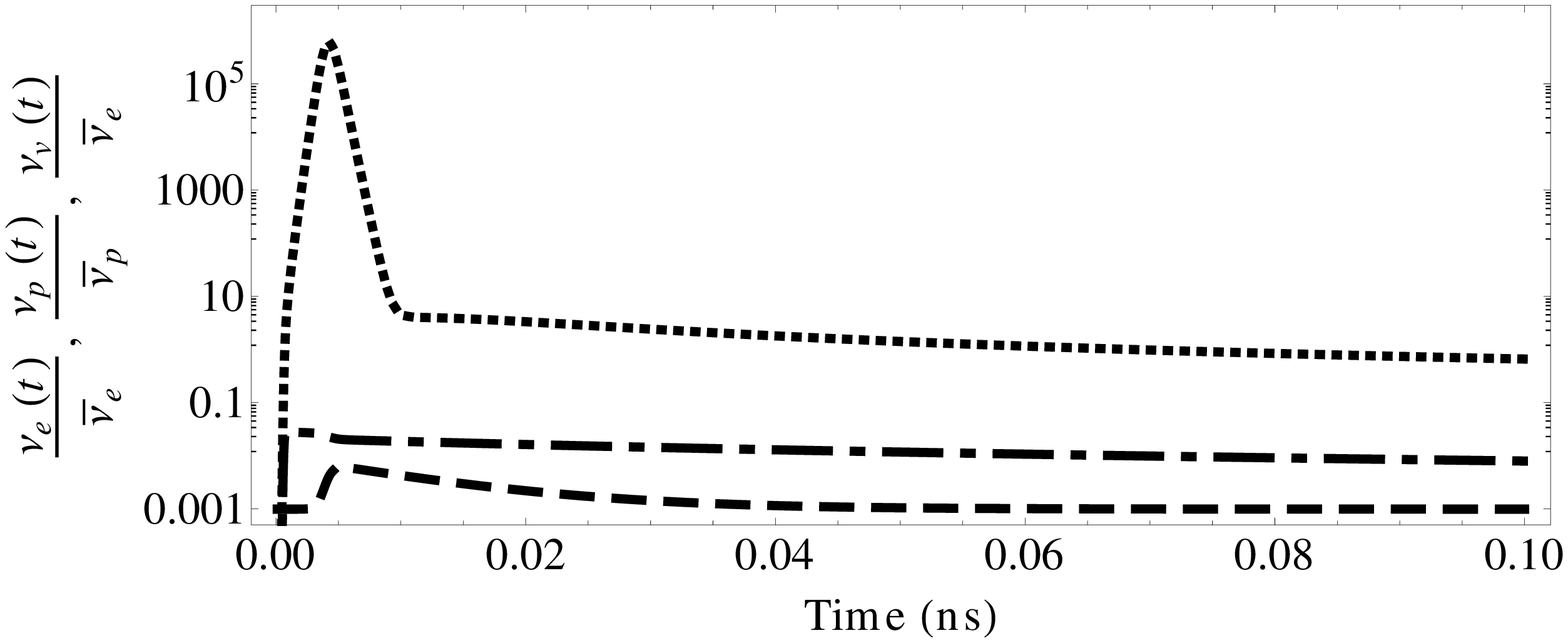}

\includegraphics[width=0.9\columnwidth]{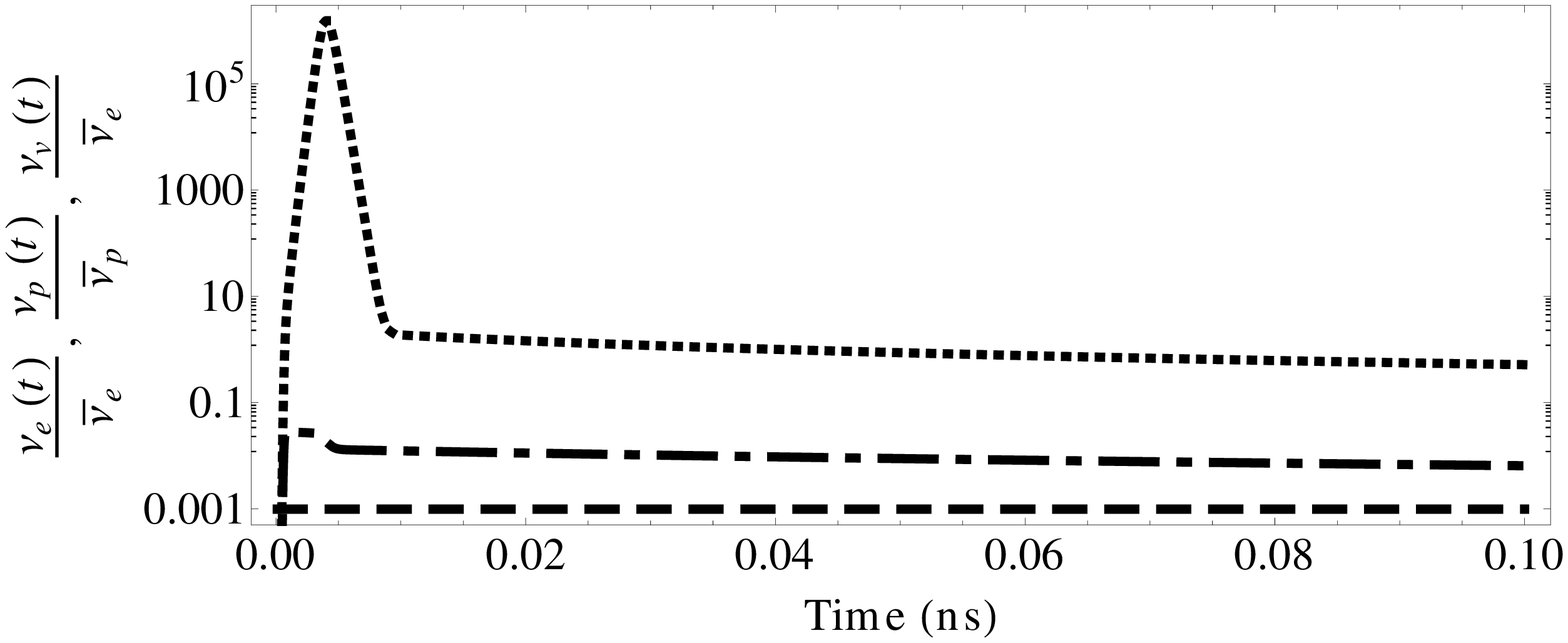}
\caption{Time-dependence of the relative surface density of exciton $\nu_e(t) / \bar \nu_e$ (dash-dotted lines), of lasing polaritons $\nu_p (t) / \bar \nu_p$ (dotted lines) and of vibrationally excited molecules $\nu_v(t) / \bar \nu_e$ (dashed lines) at threshold ($E_{\rm tot} = 300  $ nJ, $\tau_p = 85$ fs). Top: $\nu_v(t)$ is a dynamical quantity; bottom: $\nu_v(t) = \bar \nu_e / (e^{\beta E_{01}}-1)$ is static.}
\label{fig:App_S2}
\end{figure}

In Fig.~\ref{fig:App_S} we show the results, which are obtained with the same parameters used in the main text. No qualitative difference with Fig.~\ref{fig:Simple2} is observable and this refinement can not take fix the above-threshold discrepancy. 
For $\tau_p = 1 $ ps (solid line), the situation in which the back-scattering efficiency is most comparable to the polariton PL rate, a slight shift of the threshold towards higher pump fluences is observable.

Direct inspection of the time dependence of $\nu_v(t)$ shows that at threshold  the system is driven out of equilibrium on the time-scale of $5 \sim 50$ ps (Fig.~\ref{fig:App_S2}). 
However, even when $\nu_v(t)$ is consistently driven out of equilibrium, no significant difference in the population of excitons and of polaritons is observable. 
Polaritons are so few that only massive back-scattering to exciton states can affect the reservoir population. 
On the other hand, the polariton escape through the mirrors remains the dominant timescale for the polariton depletion.

\end{document}